\documentclass[conference,a4paper,10pt]{IEEEtran}
\IEEEoverridecommandlockouts
\usepackage{xcolor}
\usepackage{balance}
\usepackage{subcaption}
\usepackage{amsmath}
\usepackage{amsmath,amssymb,amstext}
\usepackage{bm}
\usepackage{algorithm}
\usepackage{algpseudocode}
\usepackage{comment}
\usepackage{caption} 
\usepackage{cite}
\usepackage{booktabs}
\usepackage{eso-pic}

\newcommand{\arxivnotice}{%
\AddToShipoutPictureFG*{%
  \AtPageLowerLeft{%
    \raisebox{25mm}{%
      \makebox[\paperwidth][c]{%
        \parbox{0.85\paperwidth}{%
          \centering
          \fontsize{6}{7}\selectfont
          This work has been submitted to the IEEE for possible publication.
          Copyright may be transferred without notice, after which this
          version may no longer be accessible.
        }%
      }%
    }%
  }%
}}

\usepackage[
    indexonlyfirst, 
    section=section,
    nonumberlist,numberedsection
]{glossaries}
\makenoidxglossaries%
\newacronym{2d}{2D}{two-dimensional}
\newacronym{3d}{3D}{three-dimensional}
\newacronym{aoa}{AoA}{angle-of-arrival}
\newacronym{aod}{AoD}{angle-of-departure}
\newacronym{awgn}{AWGN}{additive white Gaussian noise}
\newacronym{bp}{BP}{belief propagation}
\newacronym{cdf}{CDF}{cumulative distribution function}
\newacronym{csi}{CSI}{channel state information}
\newacronym{crlb}{CRLB}{Cram\'er--Rao lower bound}
\newacronym{doa}{DoA}{direction-of-arrival}
\newacronym{hcrlb}{HCRLB}{hybrid Cram\'er--Rao lower bound}
\newacronym{pcrlb}{PCRLB}{posterior CRLB}
\newacronym{dm}{DM}{diffuse multipath}
\newacronym{dmc}{DMC}{diffuse multipath component}
\newacronym{efim}{EFIM}{equivalent FIM}
\newacronym{em}{EM}{expectation maximization}
\newacronym{fg}{FG}{factor graph}
\newacronym[longplural={Fisher information matrices}]{fim}{FIM}{Fisher information matrix}
\newacronym{gpu}{GPU}{graphics processing unit}
\newacronym{iot}{IoT}{Internet of Things}
\newacronym{isac}{ISAC}{integrated sensing and communications}
\newacronym{ls}{LS}{least squares}
\newacronym{los}{LoS}{line-of-sight}
\newacronym{mc}{MC}{Monte Carlo}
\newacronym{meb}{MEB}{mapping error bound}
\newacronym{simo}{SIMO}{single-input multiple-output}
\newacronym{mimo}{MIMO}{multiple-input multiple-output}
\newacronym{miso}{MISO}{multiple-input single-output}
\newacronym{ml}{ML}{maximum likelihood}
\newacronym{music}{MUSIC}{multiple signal classification}
\newacronym{omp}{OMP}{orthogonal matching pursuit}
\newacronym{ospa}{OSPA}{optimal subpattern-assignment}
\newacronym{dmimo}{D-MIMO}{distributed MIMO}
\newacronym{xlmimo}{XL-MIMO}{extremely large-scale multiple-input multiple-output}
\newacronym{mmse}{MMSE}{minimum mean square error}
\newacronym{mmwave}{mmWave}{millimeter wave}
\newacronym{mpc}{MPC}{multipath component}
\newacronym{mpslam}{MP-SLAM}{multipath-based simultaneous localization and mapping}
\newacronym{mrt}{MRT}{maximum ratio transmission}
\newacronym{mse}{MSE}{mean squared error}
\newacronym{mt}{MT}{mobile terminal}
\newacronym{mva}{MVA}{master virtual anchor}
\newacronym{nlos}{NLoS}{non-LoS}
\newacronym{nc}{NC}{noncoherent}
\newacronym{ncv}{NCV}{nearly constant velocity}
\newacronym{oeb}{OEB}{orientation error bound}
\newacronym{olos}{OLoS}{obstructed LoS}
\newacronym{ota}{OTA}{over-the-air}
\newacronym{pa}{PA}{physical anchor}
\newacronym{bs}{BS}{base station}
\newacronym{peb}{PEB}{position error bound}
\newacronym{pdf}{PDF}{probability density function}
\newacronym{pm}{PM}{physical mobile}
\newacronym{pmf}{PMF}{probability mass function}
\newacronym{pf}{PF}{potential feature}
\newacronym{pg}{PG}{path gain}
\newacronym{pl}{PL}{path loss}
\newacronym{pr}{PBR}{particle-based representation}
\newacronym{ppr}{PR}{potential ray}
\newacronym{prop}{PROP}{proposed algorithm}
\newacronym{rcs}{RCS}{radar cross section}
\newacronym{ref}{REF}{reference algorithm}
\newacronym{rf}{RF}{radio frequency}
\newacronym{rfid}{RFID}{radio frequency identification}
\newacronym{rmse}{RMSE}{root mean square error}
\newacronym{roi}{ROI}{region of interest}
\newacronym{rssi}{RSSI}{received signal strength indicator}
\newacronym{rv}{RV}{random variable}
\newacronym{s-parameter}{S-parameter}{scattering parameter}
\newacronym{sfv}{SFV}{surface feature vector}
\newacronym{siso}{SISO}{single-input single-output}
\newacronym{sir}{SIR}{sequential importance resampling}
\newacronym{sis}{SIS}{sequential importance sampling}
\newacronym{slam}{SLAM}{simultaneous localization and mapping}
\newacronym{smc}{SMC}{specular multipath component}
\newacronym{snr}{SNR}{signal-to-noise ratio}
\newacronym{spa}{SPA}{sum-product algorithm}
\newacronym{tdoa}{TDOA}{time-difference-of-arrival}
\newacronym{toa}{TOA}{time-of-arrival}
\newacronym{tosm}{TOSM}{through – open – short – match}
\newacronym{ue}{UE}{user equipment}
\newacronym{ura}{URA}{uniform rectangular array}
\newacronym{uwb}{UWB}{ultra-wideband}
\newacronym{va}{VA}{virtual anchor}
\newacronym{vm}{VM}{virtual mobile}
\newacronym{vna}{VNA}{vector network analyzer}
\newacronym{wb}{WB}{wideband}
\newacronym{wpt}{WPT}{wireless power transfer}
\newacronym{xets}{XETS}{cross exponentially tapered slot}

\newacronym{nzm}{NZM}{nonzero-mean}
\newacronym{zm}{ZM}{zero-mean}

\newacronym{sb}{SB}{single-bounce}
\newacronym{db}{DB}{double-bounce}
\newacronym{sbl}{SBL}{sparse Bayesian learning}

\ifCLASSINFOpdf
  \usepackage[pdftex]{graphicx}
\else
  \usepackage[dvips]{graphicx}
\fi

\usepackage{pgfplots}
\pgfplotsset{compat=1.18}
\usepgfplotslibrary{colorbrewer}
\usetikzlibrary{plotmarks}

\definecolor{cNoCal}{HTML}{D55E00}   
\definecolor{cML}{HTML}{0072B2}      
\definecolor{cLS}{HTML}{009E73}      
\definecolor{cBound}{HTML}{000000}   
\definecolor{cWOnly}{HTML}{CC79A7}   
\definecolor{cDrOnly}{HTML}{E69F00}  

\pgfplotsset{
  pgf2 base/.style={
    width=0.94\linewidth,   
    height=0.62\linewidth,
    enlarge x limits=0.04,
    xtick={0.001,0.003,0.01,0.03,0.1,0.3,1,3,10,30,100},
    xticklabels={$0.001$,$0.003$,$0.01$,$0.03$,$0.1$,$0.3$,$1$,$3$,$10$,$30$,$100$},
    minor xtick={0.002,0.004,0.005,0.006,0.007,0.008,0.009,
                 0.02,0.04,0.05,0.06,0.07,0.08,0.09,
                 0.2,0.4,0.5,0.6,0.7,0.8,0.9,
                 2,4,5,6,7,8,9,20,40,50,60,70,80,90},
    xticklabel style={font=\footnotesize},
    yticklabel style={font=\footnotesize},
    label style={font=\small},
    xlabel={Calibration error std.\ $\sigma_\omega$},
    grid=both,
    scaled y ticks=false,
    major grid style={line width=0.3pt, draw=gray!30},
    minor grid style={line width=0.2pt, draw=gray!12},
    axis line style={line width=0.5pt},
    tick style={line width=0.5pt, color=black},
    tick align=inside,
    legend cell align=left,
    legend style={at={(0.5,1.03)}, anchor=south},
    legend style={
      font=\footnotesize,
      draw=gray!50, fill=white, fill opacity=0.9, text opacity=1,
      rounded corners=1pt,
      row sep=-1pt,
      inner xsep=3pt, inner ysep=2pt,
      /tikz/every even column/.append style={column sep=8pt},
    },
    every axis plot/.append style={line width=1.1pt, mark size=2.2pt},
    table/col sep=comma,
  },
  pgf2 nocal/.style={color=cNoCal, mark=*,        mark options={solid, fill=cNoCal}},
  pgf2 ml/.style   ={color=cML,    mark=square*,   mark options={solid, fill=cML}},
  pgf2 ls/.style   ={color=cLS,    mark=triangle*, densely dashed,
                     mark options={solid, fill=white, scale=1.25}},
  pgf2 bound/.style={color=cBound, dashdotted, line width=1.0pt, mark=none},
  pgf2 wonly/.style ={color=cWOnly,  mark=diamond, densely dotted,
                      mark options={solid, scale=1.7, line width=0.9pt}},
  pgf2 dronly/.style={color=cDrOnly, mark=pentagon*, densely dashdotted,
                      mark options={solid, fill=cDrOnly}},
}

\ifCLASSOPTIONcompsoc
 \usepackage[caption=false,font=normalsize,labelfont=sf,textfont=sf]{subfig}
\else
 \usepackage[caption=false,font=footnotesize]{subfig}
\fi

\usepackage{pgfplots}
\pgfplotsset{compat=newest}
\usepgfplotslibrary{groupplots} 
\usepackage{tikz}
\usetikzlibrary{calc}
\usetikzlibrary{positioning}
\usetikzlibrary{arrows.meta}
\usetikzlibrary{spy}

\colorlet{plot1}{cyan!30!white}
\colorlet{plot2}{red!40!white}
\colorlet{plot3}{cyan!70!black}
\colorlet{plot4}{black}

\newcommand{\ist}{\hspace*{.3mm}}

\newcommand{\iist}{\hspace*{1mm}}

\newcommand{\nn}{\nonumber}

\renewcommand{\H}{\mathrm{H}}

\begin{document}

\title{
Sparse Bayesian Sensing for Antenna Arrays with Per-Element Gain, Phase, and Position Errors
}

\author{
\IEEEauthorblockN{Patrick Hödl, Erik Leitinger, Klaus Witrisal}
\IEEEauthorblockA{
Institute of Comm. Networks and Satellite Comms., Graz University of Technology, Austria 
}
}

\arxivnotice
\maketitle

\begin{abstract}
The sensing capabilities of 6G wireless communication systems rely on accurate models of phased-array antennas. However, hardware imperfections can introduce a mismatch between the assumed and the actual array response, leading to performance degradations. To prevent these, we propose a parametric, model-based \gls{sbl} channel estimation algorithm with integrated self-calibration, which accounts for antenna-position deviations as well as gain and phase offsets. The algorithm learns channel parameters such as delay and \gls{doa}, together with per-antenna complex weights and position deviations based on fast variational \gls{sbl} with dictionary refinement. We derive a marginal-likelihood-based cost function for the antenna position update. The sensing performance is evaluated in terms of the \gls{ospa} metric of the delay and \gls{doa}. The proposed method approaches a benchmark based on the derived hybrid Cramér–Rao lower bound (HCRLB) in the considered single-snapshot sensing scenarios. Finally, we investigate the influence of the detection threshold using synthetic and measurement data. On single-snapshot measurements, our algorithm reduces the position OSPA from about 6~m to 0.2~m for an appropriately chosen threshold.
\end{abstract}

\glsresetall

\section{Introduction}
Array-based sensing is widely used in applications such as radar, sonar, acoustic source localization, and channel estimation for wireless communications, where \gls{doa} and delay are estimated from incoming signals. With the emergence of 6G wireless systems, array-based sensing is becoming a key technology for \gls{isac}. The accuracy and performance of such sensing systems depend on the knowledge of their current array response. Even a pre-calibrated array can deviate from its nominal characterization due to hardware imperfections and changing operating conditions. This motivates integrating self-calibration into antenna array sensing systems. Self-calibration refers to the learning of the current hardware impairments only from measurement data, without relying on dedicated reference sources or labeled calibration data. Ideally, this process should be accurate, computationally efficient, and capable of operating in situ on the sensing device, enabling adaptation to changing hardware conditions without interrupting normal operation.

\subsection{State of the Art}
Hardware deviations are known to distort the array beampattern, causing mismatches in parametric signal models~\cite{SwiKaiTSP1992},\cite{CheLiPouHuaGePetTWC2026}. The main consequence for channel parameter estimation is the detection of additional spurious components, which degrade the sensing performance \cite{LanKasReiTAP2012}, \cite{SchSchFegSteTAP2013}. Common antenna array calibration models account for complex gain impairments, effective antenna position deviations, and mutual coupling. Methods that jointly estimate these calibration parameters and the channel parameters can broadly be classified into model-based, learning-based, and hybrid approaches. Model-based approaches extend conventional estimation methods, such as \gls{ml} or \gls{music}, to incorporate self-calibration. These variants have predominantly focused on narrowband signal models, deterministic calibration parameters, and \gls{doa} estimation \cite{WeiFriTASSP1989}, \cite{BaiLiuZhangTFECS2022}. Model-based \gls{sbl} offers advantages, including super-resolution, model-order estimation, and the ability to handle coherent sources \cite{FauTip2001, WipRao:TSP2004, WipfRaoNagrajan:TIT2011,GreLeiWitFle:TWC2024}.  
In contrast, supervised learning-based methods require substantial amounts of labeled training data \cite{RamHagFurMagWym:TWC2025_HMLISAC}. More recently, there is a trend to use hybrid variants, where parts of the physical model are learned by gradient-based optimization in an unsupervised fashion. For example, \cite{ChaMarCorICC2025} reformulates \gls{music} in a differentiable form to enable gradient-based optimization.  In \cite{MatChaMagShlWymHae2026}, a learning-based approach is developed to optimize sensing performance in the presence of antenna-position and complex gain deviations, using a loss function based on the residual of the \gls{omp} algorithm. Despite avoiding the need for labeled training data, such hybrid model-based learning methods still rely on training with thousands of snapshots. This motivates self-calibration methods that exploit the signal model directly.

\subsection{Contributions}

We propose an \gls{sbl}-based approach. Instead of current \gls{sbl} versions like \cite{LiZha:PiEMR2022, CheMaCheCao:IEEE2020, WanGuiZhaFuLiu:RS2025}, which account only for weight deviations under a narrowband signal model, we integrate position deviation estimation into the fast variational \gls{sbl} algorithm in \cite{HoeMoeLeiWitICC2026} and exploit the wideband model. To validate our results, we derive the \gls{hcrlb} for the probabilistic signal model \cite{MesWSAMP2006}. Unlike data-driven, learning-based methods, our algorithm does not rely on training samples and works for a single snapshot. The main contributions are as follows.
 \begin{itemize}
     \item We extend the fast variational \gls{sbl} algorithm in \cite{HoeMoeLeiWitICC2026} by incorporating antenna-position deviations as deterministic unknown parameters and derive a marginal-likelihood-based position update that accounts for the uncertainty in the complex path amplitudes.
     \item We derive the \gls{hcrlb} for \gls{doa} and delay estimation under the considered wideband signal model with hardware impairments as a sensing performance benchmark.
    \item We show that the proposed algorithm approaches the \gls{hcrlb} benchmark in the considered scenarios when estimating coherent multipath components and outperforms weight-only and position-only calibration for larger hardware impairments.
     \item We validate the approach on synthetic data and demonstrate single-snapshot self-calibration on elementspace-transformed measurements from a 28-GHz analog phased-array frontend.
 \end{itemize}

\section{Signal Model}
We adopt the static wideband \gls{simo} setup from \cite{HoeMoeLeiWitICC2026}, consisting of a transmit antenna and a $P$-element uniform linear array (ULA) at the receiver. We extend the model to account for effective position deviations along the array axis, such that $\rho_p=\rho_p^0+\Delta\rho_p$, where $\rho_p^0$ is the nominal position of element $p \in \{1, \dots, P\}$. This yields the frequency-domain received signal model 
\begin{align}
    \bm{y}_p
    = \sum_{k=1}^{K} \alpha_k w_p a_p(\varphi_k, \Delta\rho_p) \mathrm{diag}(\bm{s}_{\text{f}}) \bm{a}_\tau(\tau_k) + \bm{n}_p \quad \in \mathbb{C}^{N \times 1}
    \label{eq:y_f_antenna}
\end{align}
for $K$ incoming signals with complex amplitudes $\alpha_k$, which is synchronously and uniformly sampled with frequency spacing $\Delta$ over the bandwidth $B$ to collect a total of $N = B/\Delta$ samples of the Fourier spectrum $S(f)$ of the transmitted signal $s(t)$ in the vector  $\bm{s}_{\mathrm{f}} = \bigl[\, S(-\tfrac{N}{2}\Delta) \;\cdots\; S((\tfrac{N}{2}-1)\Delta) \,\bigr]^{\mathrm{T}}$. The temporal response of the array is described by $\bm{a}_\tau(\tau_k) = \exp\!\big(-j 2\pi \bm{f} \tau_k\big)\in \mathbb{C}^{N\times 1}$, with $\bm{f} = \bigl[\, -\tfrac{N}{2}\Delta \;\cdots\; (\tfrac{N}{2}-1 )\Delta \,\bigr]^{\mathrm{T}}$ holding the uniformly spaced baseband frequency points and $\tau_k$ describing the delay of the signal component $k$. The spatial response $a_p(\varphi_k,\Delta \rho_p) = \exp\!\big(-\mathrm{j} \tfrac{2\pi}{\lambda_\text{c}} (\rho_p^0+\Delta\rho_p) \sin\varphi_k\big)$ of an array element depends on the \gls{doa} of the incoming wave $\varphi_k$ w.r.t.\ broadside, the carrier wavelength $\lambda_\text{c} = c/f_\text{c}$, and the perturbed element position $\rho_p^0+\Delta\rho_p$ along the array axis. The per-element gain and phase errors are modeled as a complex weight  $w_p$, and $n_p$ denotes the measurement noise vector given by a complex circularly symmetric Gaussian random vector with covariance matrix $\lambda^{-1}\bm{I}_{N}$, where $\lambda$ is the noise precision.

To get a compact generative model we stack the signals $\bm{y}_p$ into $\bm{y}=[\bm{y}_1^{\mathrm{T}}\cdots\bm{y}_P^{\mathrm{T}}]^{\mathrm{T}}\in\mathbb{C}^{PN\times 1}$ and express it as
\begin{equation}
    \bm{y} = \bm{D}(\bm{w})\bm{A}(\bm{\theta} ,\Delta\bm{\rho})\bm{\alpha} + \bm{n}
    \label{eq:y_compact_final}
\end{equation}
where $\bm{\theta} = [\bm\varphi^{\mathrm{T}},\bm\tau^{\mathrm{T}}]^{\mathrm{T}}$, $\bm{\alpha}=[\alpha_1 \cdots\alpha_K]^{\mathrm{T}}$, $\bm{D}(\bm{w})=\mathrm{diag}(\bm{w}\otimes\bm{1}_N)$ $\in \mathbb{C}^{PN \times PN}$ and the dictionary matrix 
\begin{equation}
\begin{aligned}
    \bm{A}(\bm{\theta}, \Delta\bm{\rho})
    &=
    \big[
    \bm{a}(\varphi_1, \Delta\bm{\rho})
    \otimes
    \big(\mathrm{diag}(\bm{s}_{\text{f}})\bm{a}_\tau(\tau_1)\big)
    \cdots
    \\
    &\quad
    \bm{a}(\varphi_K, \Delta\bm{\rho})
    \otimes
    \big(\mathrm{diag}(\bm{s}_{\text{f}})\bm{a}_\tau(\tau_K)\big)
    \big]
    \in \mathbb{C}^{PN \times K}
\end{aligned}
\end{equation}
 with $\bm{a}(\varphi_k,\Delta\bm{\rho}) = [a_1(\varphi_k,\Delta\rho_1) \cdots a_P(\varphi_k,\Delta\rho_P)]^\text{T}$ and $\otimes$ denoting the Kronecker-product \cite{RichterPhD2005}. The stacked noise vector $\bm{n} = [\bm{n}_1^{\mathrm{T}} \cdots \bm{n}_P^{\mathrm{T}}]^{\mathrm{T}}$ follows a circularly symmetric complex Gaussian distribution with covariance matrix $\lambda^{-1}\bm{I}_{PN}$.

\section{Bayesian Estimation}
\subsection{Probabilistic Model}
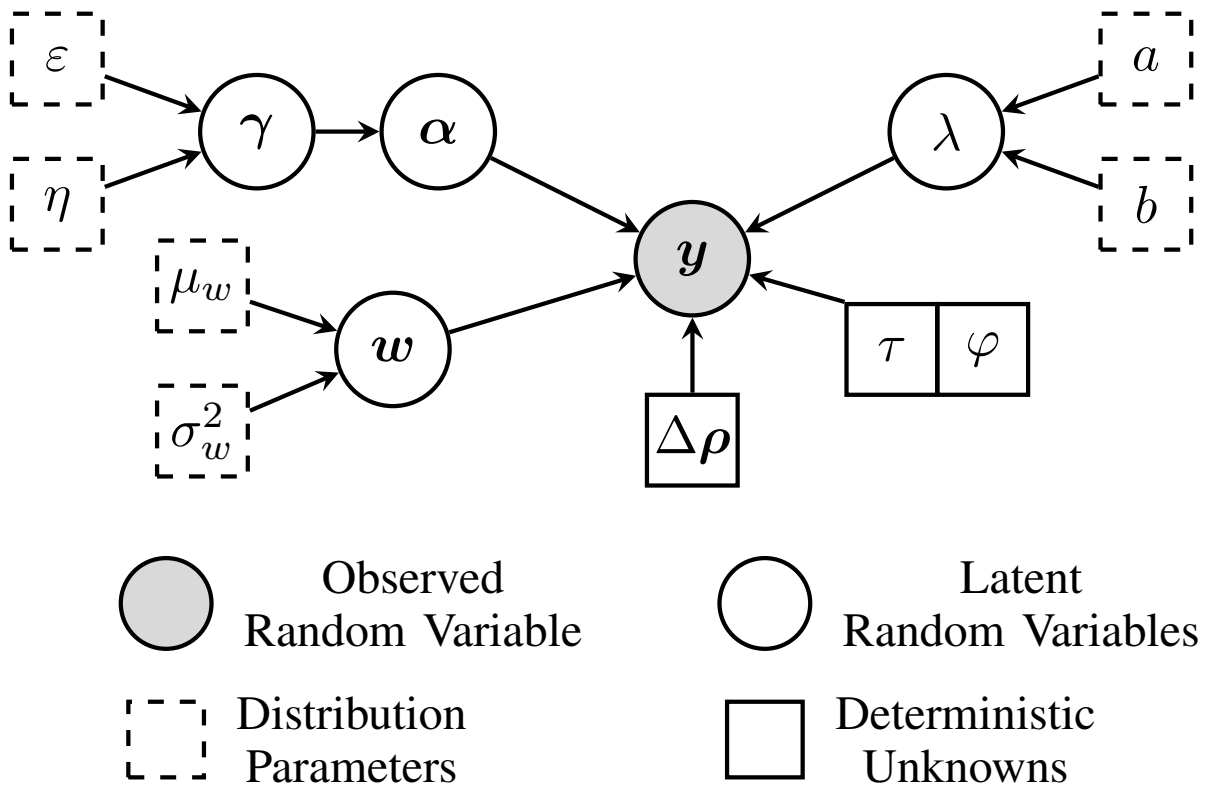
\begin{figure}[!t]
  \centering
  \begin{tikzpicture}[
  scale=1.2,
  >=stealth, thick,
  hid/.style   ={circle, draw, minimum size=7.5mm, inner sep=0pt},
  obs/.style   ={hid, fill=gray!30},
  par/.style   ={rectangle, draw, dashed, minimum size=6mm, inner sep=1pt},
  det/.style   ={rectangle, draw, minimum size=6mm, inner sep=1pt},
  lbl/.style   ={font=\small, align=center}
]
 
\node[hid] (gamma)  at (1.0, 1.6) {$\bm{\gamma}$};
\node[hid] (alpha)  at (2.0, 1.6) {$\bm{\alpha}$};
\node[hid] (lambda) at (4.8, 1.6) {$\lambda$};
\node[hid] (omega)  at (1.75, 0.4) {$\bm{w}$};
\node[obs] (y)      at (3.4, 0.9) {$\bm{y}$};
 
\node[par] (eps)   at (-0.1, 2.0) {$\varepsilon$};
\node[par] (eta)   at (-0.1, 1.2) {$\eta$};
\node[par] (a)     at (5.9, 2.0) {$a$};
\node[par] (b)     at (5.9, 1.2) {$b$};
\node[par] (muw)   at (0.7, 0.75) {$\mu_w$};
\node[par] (sigw)  at (0.7,-0.05) {$\sigma_w^2$};
 
\node[det] (drho)  at (3.4,-0.1) {$\Delta \bm{\rho}$};
\node[det] (tau)   at (4.5, 0.4) {$\tau$};
\node[det, right=-\pgflinewidth of tau] (phi) {$\varphi$};
 
\draw[->] (eps)    -- (gamma);
\draw[->] (eta)    -- (gamma);
\draw[->] (gamma)  -- (alpha);
\draw[->] (alpha)  -- (y);
\draw[->] (a)      -- (lambda);
\draw[->] (b)      -- (lambda);
\draw[->] (lambda) -- (y);
\draw[->] (muw)    -- (omega);
\draw[->] (sigw)   -- (omega);
\draw[->] (omega)  -- (y);
\draw[->] (drho)   -- (y);
\draw[->] (tau.north west) -- (y);
 
\begin{scope}[yshift=-1.0cm]
  \node[obs, minimum size=6mm] (l1) at (0.5, 0) {};
  \node[lbl, right=2pt of l1] {Observed\\Random Variable};
  \node[hid, minimum size=6mm] (l2) at (3.8, 0) {};
  \node[lbl, right=2pt of l2] {Latent\\Random Variables};
  \node[par, minimum size=5mm] (l3) at (0.5,-0.75) {};
  \node[lbl, right=2pt of l3] {Distribution\\Parameters};
  \node[det, minimum size=5mm] (l4) at (3.8,-0.75) {};
  \node[lbl, right=2pt of l4] {Deterministic\\Unknowns};
\end{scope}
 
\end{tikzpicture}
  \caption{Bayesian hierarchical Gamma-Gaussian prior model.}
  \label{fig:GraphBayes}
\end{figure}

We adopt the sparsity-inducing hierarchical Bayesian Gamma–Gaussian prior model from \cite{HoeMoeLeiWitICC2026} and extend it to include the position deviations $\Delta\bm{\rho}$, as illustrated in Fig.~\ref{fig:GraphBayes}.
 Since $\Delta\bm{\rho}$ enters the steering vectors nonlinearly, a Gaussian prior
  would not yield closed-form variational updates without further approximations. Hence, analogous to the channel parameters $\bm{\theta}$
  in~\cite{HoeMoeLeiWitICC2026}, we model $\Delta\bm{\rho}$ as deterministic unknowns. The goal is to estimate the posterior moments of the latent random variables and the deterministic unknowns from the observed vector $\bm{y}$. The chosen probability distributions are
\vspace*{-1mm}
\begin{align}
    p(\bm{\alpha}|\bm{\gamma})
	= \prod_{k=1}^{K_{\text{max}}} \mathrm{CN}\big(\alpha_k\big|0,\gamma_k^{-1}\big)
	\label{eq:prior_alpha}\\[-6mm]\nn
\end{align}
where $K_\mathrm{max}\geq K$ includes virtual components with $\alpha_k=0$,  which are driven to zero by the sparsity-promoting prior, so that $K$ need not be
  known a priori. Each precision parameter $\gamma_k$ is given by
$p(\gamma_k)=\mathrm{Ga}(\gamma_k\mid\epsilon,\eta)$, where $\mathrm{Ga}(\cdot\mid\epsilon,\eta)$ is a Gamma \gls{pdf} with shape parameter $\epsilon$ and rate $\eta$, chosen as the Jeffreys prior in the limit $\epsilon,\eta\to 0$.
The complex weight deviations are modeled as
\vspace*{-1mm}
\begin{align}
    p(\bm{w}) = \prod_{p=1}^{P} \mathrm{CN} \big(w_p\big|\mu_{\text{w},p},\sigma_{\text{w},p}^2 \big)
	\label{eq:prior_w}\\[-6mm]\nn
\end{align}
with means $\mu_{\text{w},p}=1$ and variances $\sigma_{\text{w},p}^2=100$, yielding a weakly informative prior. The noise precision $\lambda$ is modeled as an independent Gamma-distributed random variable with PDF $p(\lambda) = \mathrm{Ga}(\lambda| a, b)$, again as the Jeffreys prior with $a,b\to 0$. This results in the likelihood of the observed signal in \eqref{eq:y_compact_final} given by
\begin{equation}
\begin{aligned}
    p(\bm{y}|\bm{w},\bm{\alpha},\lambda;
    \bm{\tau},\bm{\varphi},\Delta\bm{\rho})
    &= 
    \mathrm{CN}\Big(
    \bm{y}\,\Big|\,
    \bm{D}(\bm{w})
    \bm{A}(\bm{\theta},\Delta\bm{\rho})
    \bm{\alpha},
    \lambda^{-1}\bm{I}
    \Big)
\end{aligned}
\label{eq:likelihood}
\end{equation}
Based on this introduced model, the corresponding joint posterior PDF is proportional to
\begin{align}
    &p(\bm{w},\bm{\alpha},\bm{\gamma},\lambda|\bm{y};\bm{\theta},\Delta\bm{\rho}) \notag  \\
    & \hspace{12mm} \propto
    p(\bm{y}|\bm{w},\bm{\alpha},\lambda;\bm{\theta},\Delta\bm{\rho})
    p(\bm{w})
    p(\bm{\alpha}|\bm{\gamma})
    p(\bm{\gamma})
    p(\lambda).
    \label{eq:joint}
\end{align}
\subsection{Variational Updates}

We apply variational Bayesian inference to approximate the posterior \gls{pdf} in \eqref{eq:joint} \cite{TziLikGal:SPM2008, moederlFusion2025:multi-dictionary-SBL,MoePerWitLei:TSP2024}. We explicitly derive the update for the antenna position deviations $\Delta\bm{\rho}$ and adopt the remaining variational updates from \cite{HoeMoeLeiWitICC2026}. The proposed variational \gls{em} algorithm alternates between (i) updating approximate posterior distributions of $\bm{w}$, $\bm{\alpha}$, and $\lambda$ in the E-step and (ii) applying fast updates for $\bm{\gamma}$ and the deterministic parameters $\bm{\theta}$ as well as the position deviations $\Delta\bm{\rho}$ in the M-step. Since we model the position deviations as deterministic parameters, incorporating their point estimates only requires updating the dictionary matrix as $\hat{\bm{A}} \triangleq \bm{A}(\hat{\bm{\theta}}, \Delta\hat{\rho})$. The update rules provided in \cite{HoeMoeLeiWitICC2026} therefore remain unchanged, apart from using the updated dictionary matrix.

\subsection{Marginal Likelihood Update for Position Deviations}
In the following, $\hat{(\cdot)}$ denotes the current estimate, i.e., the posterior mean for random variables and the point estimate for deterministic parameters. To remain consistent with the existing \gls{sbl} framework, we incorporate the uncertainty of the latent complex amplitudes $\bm{\alpha}$ in our cost function. We start with the likelihood function in \eqref{eq:likelihood}, conditioned on the complex amplitudes and parametrized by the estimates of the other unknowns, and marginalize out $\bm{\alpha}$ as 
\begin{equation}
\begin{aligned}
    L(\Delta\bm{\rho} )
    &=
    \int
    p\!\left(
        \bm y
        \middle|
        \bm\alpha;\Delta\bm \rho,
        \hat{\bm\theta},\hat{\lambda},\hat{\bm w}
    \right)
    p\!\left(
        \bm\alpha\middle|\hat{\bm\gamma}
    \right)
    \,\mathrm d\bm\alpha 
\end{aligned}
\label{eq:position_marginal_likelihood_integral}
\end{equation}
which results in 
\begin{equation}
    L(\Delta\bm \rho)
    =
    \mathcal{CN}
    \left(
        \bm y
        \middle|
        \bm 0,
        \bm C_y(\Delta\bm \rho)
    \right)
\end{equation}
where
\begin{equation}
    \bm C_y(\Delta\bm \rho)
    =
    \hat{\lambda}^{-1}\bm I
    +
     \tilde{\bm A}(\Delta\bm\rho)
    \hat{\bm\Gamma}^{-1}
   \tilde{\bm A}^{\mathrm H}(\Delta\bm \rho)
\label{eq:position_Cy}
\end{equation}
with
\begin{equation}
  \tilde{\bm A}(\Delta\bm\rho)
  =
  \bm D(\hat{\bm w})\,
  \bm A(\hat{\bm\theta},\Delta\bm\rho)\iist\iist\iist \text{and}
  \iist\iist\iist
  \hat{\bm\Gamma} = \mathrm{diag}(\hat{\bm\gamma}).
\end{equation}

The negative log-marginal likelihood is given by
\begin{equation}
    \ell(\Delta\bm\rho)
    \triangleq
    -\ln L(\Delta\bm\rho)
    \overset{e}{\propto}
    \ln\left|\bm C_y(\Delta\bm\rho)\right|
    +
    \bm y^{\mathrm H}\bm C_y^{-1}(\Delta\bm\rho)\bm y,
\label{eq:position_log_marginal}
\end{equation}
where $\overset{e}{\propto}$ denotes equality up to an additive constant. The matrix determinant lemma and the Woodbury identity yield the equivalent form
\begin{equation}
    \ell(\Delta\bm\rho)
    \overset{e}{\propto}
    -\ln|\hat{\bm\Sigma}_{\alpha}(\Delta\bm\rho)|
    -
    \hat{\bm{\alpha}}^{\mathrm H}(\Delta\bm\rho)
    \hat{\bm\Sigma}_{\alpha}^{-1}(\Delta\bm\rho)
    \hat{\bm{\alpha}}(\Delta\bm\rho),
\label{eq:app_position_objective_mu}
\end{equation}
where $\hat{\bm{\alpha}}(\Delta\bm\rho)$ and $\hat{\bm\Sigma}_{\alpha}(\Delta\bm\rho)$ denote the posterior mean and covariance of $\bm\alpha$, defined as in the update rule~\cite[Eq.~(16)]{HoeMoeLeiWitICC2026} with $\bm A(\hat{\bm\theta},\Delta\bm\rho)$ in place of $\hat{\bm A}$. Point estimates of the position deviations are thus obtained as
\begin{equation}
    \Delta\hat{\bm\rho}
    =
    \arg\min_{\Delta\bm\rho\in\mathbb R^P}
    \ell(\Delta\bm\rho).
\label{eq:app_position_update_final}
\end{equation}

The cost is minimized element-wise for each $p$ via Brent's bounded method over
  $|\Delta\rho_p| \le 0.5\lambda_\mathrm{c}$, once per iteration.

\subsection{Implementation}
To promote fast convergence, we integrate the estimation of $\Delta\bm\rho$ into the
self-calibration step of the algorithm in~\cite[Alg.~1, line~14]{HoeMoeLeiWitICC2026}. Starting from an empty model, each iteration first performs a \emph{detection} step, where the component with the highest estimated \gls{snr} is initialized via a grid search over delay
and \gls{doa} on the observation $\bm y$, after subtracting the
already detected components. It is added only if this \gls{snr} exceeds the detection threshold $\chi$. In the subsequent \emph{self-calibration} step, the complex weights $\hat{\bm w}$ and the position deviations $\Delta\hat{\bm\rho}$ are updated given the current channel estimates. The \emph{component update} then refines the parameters
$\hat{\bm\theta}_k$ and precisions $\hat\gamma_k$ of all detected
components and prunes those whose estimated \gls{snr} falls
below $\chi$, before the amplitudes $\hat{\bm\alpha}$ and the
noise precision $\hat\lambda$ are updated.
These steps are iterated until the channel and calibration
parameters converge.

\section{Results}
We evaluate the sensing performance of our algorithm using both synthetic data generated according to the signal model in \eqref{eq:y_compact_final} and measurement data acquired from an \gls{uwb} channel sounder with an analog uniform linear phased-array receiver frontend \cite{HoeWitEuCAP2025}. The antenna elements are nominally spaced at $\lambda_c/2$. The sensing signal covers a bandwidth of $B=2$~GHz and is upconverted to a carrier frequency of $f_c=28$~GHz.

\subsection{Synthetic Data}
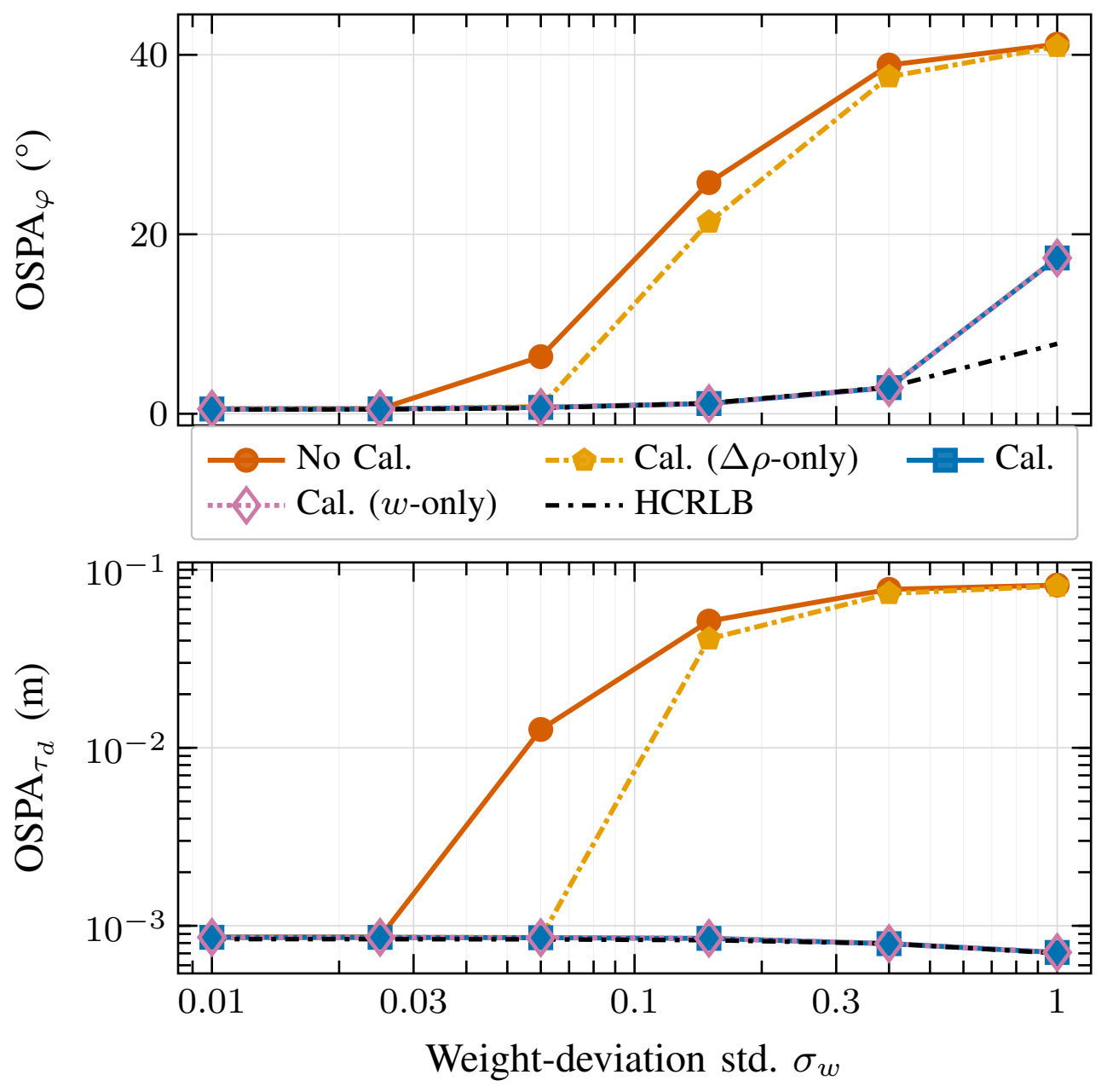
\begin{figure}[]
  \centering
%
\def\ospagap{0pt}%
\def\ospalegh{0.12\linewidth}%
\ifdefined\ospadown\else\newlength{\ospadown}\fi
\setlength{\ospadown}{\dimexpr\ospalegh+\ospagap+\ospagap\relax}%
\hspace*{0.08\linewidth}%
{%
\tikzset{every picture/.append style={trim axis left, trim axis right}}%
\begin{tikzpicture}
\pgfplotsset{ospa panel/.style={
  pgf2 base,
  scale only axis, width=0.80\linewidth, height=0.36\linewidth,
  yticklabel style={text width=2.0em, align=right, inner xsep=2pt},
  xlabel={Weight-deviation std. $\sigma_w$},
  ylabel style={at={(ticklabel* cs:0.5, 2.4em)}, anchor=south},
  enlarge y limits=false,
  yminorgrids=false,
  legend columns=3,
}}
\begin{axis}[
  ospa panel,
  xmode=log,
  xlabel={}, xticklabels={},
  ymin=-1.3, ymax=44.5, ytick={0,20,40},
  ylabel={OSPA$_{\varphi}$ (°)},
  legend style={at={(0.5,0)}, anchor=north, yshift=-\ospagap},
]
  \addplot+[pgf2 nocal] table[x=sigma_sim, y=ospa_phi_fvsbl_m2_no_cal_m_1]
    {pgf2/compare_ospa_phi_light.csv};
  \addlegendentry{No Cal.}

  \addplot+[pgf2 dronly] table[x=sigma_sim, y=ospa_phi_fvsbl_m2_delta_r_only_cal_m_1_marginal_likelihood]
    {pgf2/compare_ospa_phi_light.csv};
  \addlegendentry{Cal.\ ($\Delta \rho$-only)}

  \addplot+[pgf2 ml] table[x=sigma_sim, y=ospa_phi_fvsbl_m2_cal_m_1_marginal_likelihood]
    {pgf2/compare_ospa_phi_light.csv};
  \addlegendentry{Cal.}

  \addplot+[pgf2 wonly] table[x=sigma_sim, y=ospa_phi_fvsbl_m2_w_only_cal_m_1]
    {pgf2/compare_ospa_phi_light.csv};
  \addlegendentry{Cal.\ ($w$-only)}

  \addplot+[pgf2 bound] table[x=sigma_sim, y=ospa_phi_hcrb_model_2_position_error_averaged_over_priors_wideband_m_1]
    {pgf2/compare_ospa_phi_light.csv};
  \addlegendentry{HCRLB}

  \coordinate (ospatopbot) at (rel axis cs:0.5,0);
\end{axis}
\begin{axis}[
  ospa panel,
  at={(ospatopbot)}, anchor=north, yshift=-\ospadown,
  xmode=log,
  ymode=log,
  ymin=5.4e-4, ymax=1.1e-1,
  ylabel={OSPA$_{\tau_d}$ (m)},
  legend to name=ospaLegendLightHidden,
]
  \addplot+[pgf2 nocal] table[x=sigma_sim, y=ospa_tau_fvsbl_m2_no_cal_m_1]
    {pgf2/compare_ospa_tau_light.csv};
  \addlegendentry{No Cal.}

  \addplot+[pgf2 dronly] table[x=sigma_sim, y=ospa_tau_fvsbl_m2_delta_r_only_cal_m_1_marginal_likelihood]
    {pgf2/compare_ospa_tau_light.csv};
  \addlegendentry{Cal.\ ($\Delta \rho$-only)}

  \addplot+[pgf2 ml] table[x=sigma_sim, y=ospa_tau_fvsbl_m2_cal_m_1_marginal_likelihood]
    {pgf2/compare_ospa_tau_light.csv};
  \addlegendentry{Cal.}

  \addplot+[pgf2 wonly] table[x=sigma_sim, y=ospa_tau_fvsbl_m2_w_only_cal_m_1]
    {pgf2/compare_ospa_tau_light.csv};
  \addlegendentry{Cal.\ ($w$-only)}

  \addplot+[pgf2 bound] table[x=sigma_sim, y=ospa_tau_hcrb_model_2_position_error_averaged_over_priors_wideband_m_1]
    {pgf2/compare_ospa_tau_light.csv};
  \addlegendentry{HCRLB}
\end{axis}
\end{tikzpicture}%
}%
  \caption{Influence of the hardware impairments: \gls{rmse} of the \gls{ospa} metric of the \gls{doa} estimates (top) and of the delay estimates
    (bottom) versus $\sigma_w$, $\sigma_{\Delta \rho}=0.1\lambda_c$.}
  \label{fig:ospa_light}
\end{figure}

\begin{figure}[]
  \centering
%
\def\ospagap{0pt}%
\def\ospalegh{0.12\linewidth}%
\ifdefined\ospadown\else\newlength{\ospadown}\fi
\setlength{\ospadown}{\dimexpr\ospalegh+\ospagap+\ospagap\relax}%
\hspace*{0.08\linewidth}%
{%
\tikzset{every picture/.append style={trim axis left, trim axis right}}%
\begin{tikzpicture}
\pgfplotsset{ospa panel/.style={
  pgf2 base,
  scale only axis, width=0.80\linewidth, height=0.36\linewidth,
  yticklabel style={text width=2.0em, align=right, inner xsep=2pt},
  xlabel={Weight-deviation std. $\sigma_w$},
  ylabel style={at={(ticklabel* cs:0.5, 2.4em)}, anchor=south},
  enlarge y limits=false,
  yminorgrids=false,
  legend columns=3,
}}
\begin{axis}[
  ospa panel,
  xmode=log,
  xlabel={}, xticklabels={},
  ymin=-1.3, ymax=44.5, ytick={0,20,40},
  ylabel={OSPA$_{\varphi}$ (°)},
  legend style={at={(0.5,0)}, anchor=north, yshift=-\ospagap},
]
  \addplot+[pgf2 nocal] table[x=sigma_sim, y=ospa_phi_fvsbl_m2_no_cal_m_1]
    {pgf2/compare_ospa_phi.csv};
  \addlegendentry{No Cal.}

  \addplot+[pgf2 dronly] table[x=sigma_sim, y=ospa_phi_fvsbl_m2_delta_r_only_cal_m_1_marginal_likelihood]
    {pgf2/compare_ospa_phi.csv};
  \addlegendentry{Cal.\ ($\Delta \rho$-only)}

  \addplot+[pgf2 ml] table[x=sigma_sim, y=ospa_phi_fvsbl_m2_cal_m_1_marginal_likelihood]
    {pgf2/compare_ospa_phi.csv};
  \addlegendentry{Cal.}

  \addplot+[pgf2 wonly] table[x=sigma_sim, y=ospa_phi_fvsbl_m2_w_only_cal_m_1]
    {pgf2/compare_ospa_phi.csv};
  \addlegendentry{Cal.\ ($w$-only)}

  \addplot+[pgf2 bound] table[x=sigma_sim, y=ospa_phi_hcrb_model_2_position_error_averaged_over_priors_wideband_m_1]
    {pgf2/compare_ospa_phi.csv};
  \addlegendentry{HCRLB}

  \coordinate (ospatopbot) at (rel axis cs:0.5,0);
\end{axis}
\begin{axis}[
  ospa panel,
  at={(ospatopbot)}, anchor=north, yshift=-\ospadown,
  xmode=log,
  ymode=log,
  ymin=5.4e-4, ymax=1.1e-1,
  ylabel={OSPA$_{\tau_d}$ (m)},
  legend to name=ospaLegendHidden,
]
  \addplot+[pgf2 nocal] table[x=sigma_sim, y=ospa_tau_fvsbl_m2_no_cal_m_1]
    {pgf2/compare_ospa_tau.csv};
  \addlegendentry{No Cal.}

  \addplot+[pgf2 dronly] table[x=sigma_sim, y=ospa_tau_fvsbl_m2_delta_r_only_cal_m_1_marginal_likelihood]
    {pgf2/compare_ospa_tau.csv};
  \addlegendentry{Cal.\ ($\Delta \rho$-only)}

  \addplot+[pgf2 ml] table[x=sigma_sim, y=ospa_tau_fvsbl_m2_cal_m_1_marginal_likelihood]
    {pgf2/compare_ospa_tau.csv};
  \addlegendentry{Cal.}

  \addplot+[pgf2 wonly] table[x=sigma_sim, y=ospa_tau_fvsbl_m2_w_only_cal_m_1]
    {pgf2/compare_ospa_tau.csv};
  \addlegendentry{Cal.\ ($w$-only)}

  \addplot+[pgf2 bound] table[x=sigma_sim, y=ospa_tau_hcrb_model_2_position_error_averaged_over_priors_wideband_m_1]
    {pgf2/compare_ospa_tau.csv};
  \addlegendentry{HCRLB}
\end{axis}
\end{tikzpicture}%
}%
  \caption{Influence of the hardware impairments: \gls{rmse} of the \gls{ospa} metric of the \gls{doa} estimates (top) and of the delay estimates
    (bottom) versus $\sigma_w$, $\sigma_{\Delta \rho}=0.3\lambda_c$.}
  \label{fig:ospa_severe}
\end{figure}
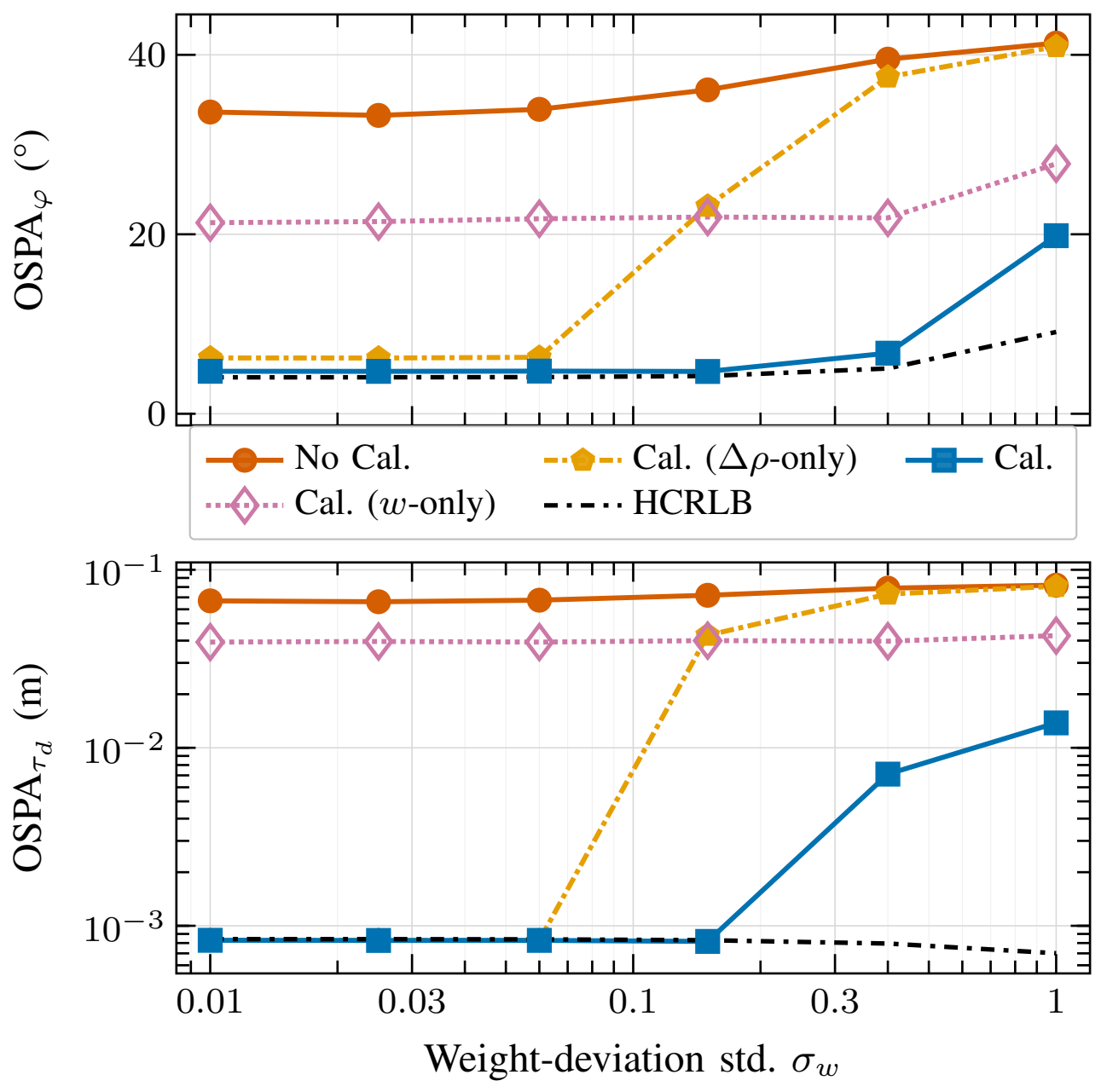

We draw our hardware deviations as $w_p \sim \mathcal{CN}(1,\sigma_{w}^2)$ and $\Delta \rho_p \sim \mathcal{N}(0,\sigma_{\Delta \rho}^2)$ and consider two simulation studies.
\newline
\newline
\textbf{Sensing performance over impairment severity:}
 First, we evaluate the sensing performance in terms of the estimated \gls{doa} $\hat{\varphi}$ and the delay expressed as an equivalent distance, $\hat{\tau}_d = c \hat{\tau}$, for various levels of hardware impairment severity. Different variants of our algorithm are compared against the \gls{hcrlb} benchmark (see the Appendix). 
 We use the \gls{ospa} metric \cite{SchVoVo:TSP2008} to jointly account for estimation and cardinality errors, with cutoff parameters of $50$° and $0.1$~m, and order $q=2$. We consider a fixed, well-conditioned scenario with $K=3$ resolvable components and sweep the standard deviation $\sigma_w$ of the complex weight deviations over $I=100$ Monte Carlo runs. In each run, the position deviations $\Delta \rho$ are drawn once and kept fixed throughout the sweep. Thus, we get $I$ different hardware deviations per $\sigma_w$ point, using the same $I$ realizations of the position deviations at every point in the sweep. We report the \gls{rmse} of the \gls{ospa} metric for delay and \gls{doa} estimation. We compare joint weight and position calibration (Cal.), weight-only calibration (Cal. ($w$-only)), position-only calibration (Cal. ($\Delta\rho$-only)), and no calibration (No Cal.). Fig. \ref{fig:ospa_light} shows the results for a low level of position deviations ($\sigma_{\Delta \rho}=0.1\lambda_c$), while Fig. \ref{fig:ospa_severe} displays the results with a higher level of position deviations ($\sigma_{\Delta \rho}=0.3\lambda_c$). Both figures show the \gls{rmse} of the \gls{ospa} metric over the Monte Carlo runs.
\newline
 Fig. \ref{fig:ospa_light} illustrates that, for low position deviations, the algorithm with weight calibration only performs comparably to the variant that jointly calibrates the complex weights and antenna positions. These variants also approach the \gls{hcrlb}, except for very high weight deviations. Applying position calibration only, though, significantly degrades the sensing performance when the standard deviation of the weight deviations exceeds $0.06$. The variant without calibration is unable to reach the \gls{hcrlb}, except for very low levels of hardware impairment. Fig. \ref{fig:ospa_severe} demonstrates the influence of larger antenna position deviations. The algorithm with weight-only calibration performs closer to the uncalibrated variant than to the \gls{hcrlb}. The position-only calibration variant performs comparably to Fig. \ref{fig:ospa_light}, but worse for weight standard deviations below $0.06$. The algorithm jointly calibrating the complex weights and antenna positions approaches the \gls{hcrlb} over a wide range of weight deviations. At very high weight-deviation standard deviations of $\sigma_w=0.3$ and $\sigma_w=1$, some Monte Carlo realizations contain particularly unfavorable hardware deviations, resulting in large estimation errors that drastically increase the \gls{ospa}. 
  \newline 
  
 \label{subsec:eval_fixed}

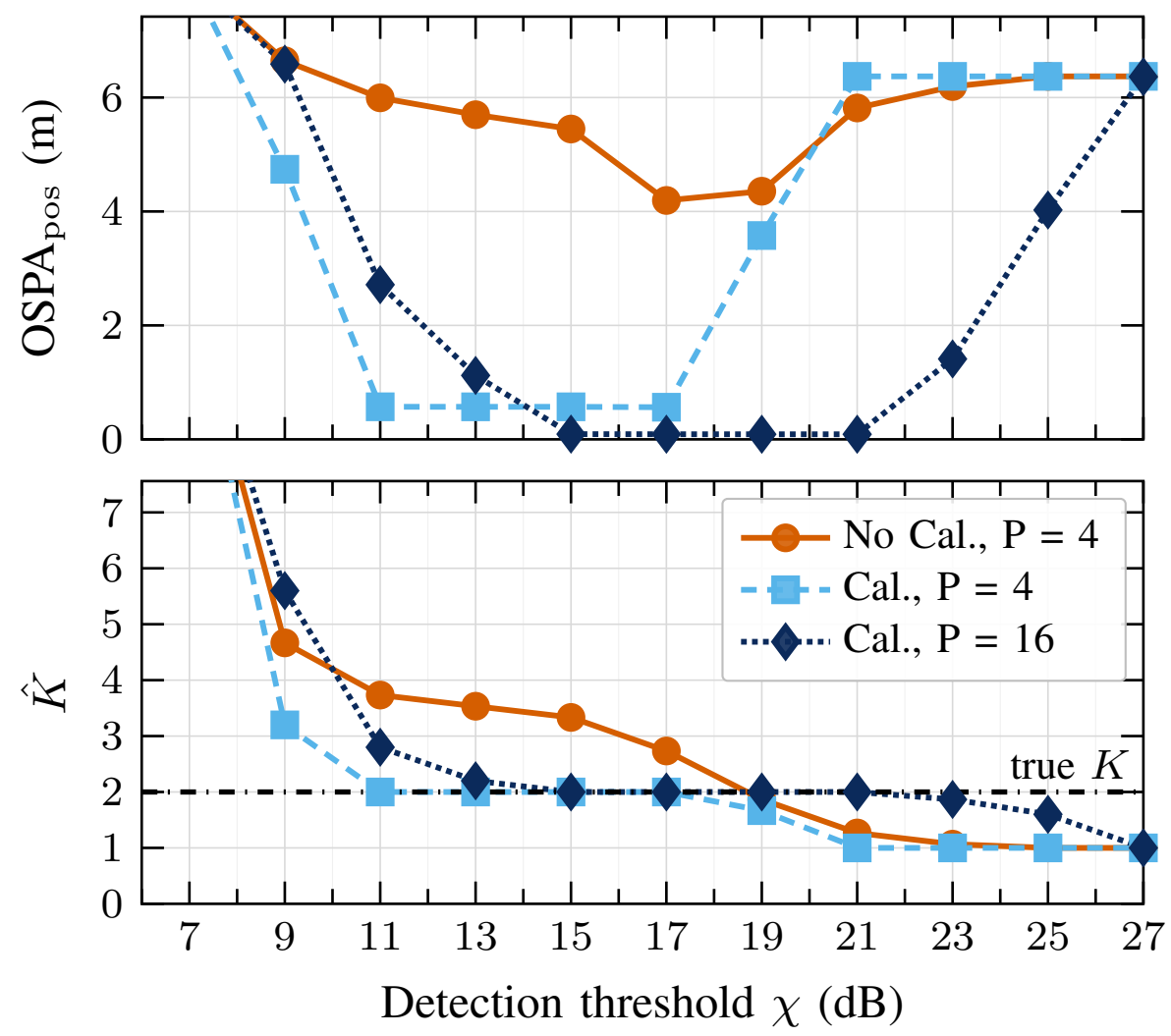
\begin{figure}[]
  \centering
%
%
\usepgfplotslibrary{groupplots}%
\def\sweepmetric{8}
\def\sweepkhat{14}
\def\chimin{6}\def\chimax{27}%
\def\pgftwoempty{}%
\def\pgftwozap#1 #2{#1\ifx#2\pgftwoempty\else\expandafter\pgftwozap\fi#2}%
\def\pgftwosweepkeep#1#2#3#4{%
  \def\pgftwosweeptest##1#1##2\pgftwosweepend{%
    \ifx\relax##2\relax\def\pgfmathresult{nan}\fi}%
  \pgftwosweeptest#3#1\pgftwosweepend
  \edef\pgftwosweepA{\pgftwozap#4 \pgftwoempty}%
  \edef\pgftwosweepB{\pgftwozap P=#2 \pgftwoempty}%
  \ifx\pgftwosweepA\pgftwosweepB\else\def\pgfmathresult{nan}\fi}%
\pgfplotsset{
  sweep table/.style={table/col sep=comma, table/header=false, table/skip first n=1},
  sweep variant/.style 2 args={
    x filter/.code={%
      \edef\pgftwosweepdo{\noexpand\pgftwosweepkeep{#1}{#2}{\thisrowno{0}}{\thisrowno{1}}}%
      \pgftwosweepdo},
  },
}
\providecolor{cP4}{HTML}{56B4E9}%
\providecolor{cP8}{HTML}{0072B2}%
\providecolor{cP16}{HTML}{0B2A5B}%
\pgfplotsset{
  sweep nocal/.style={color=cNoCal, mark=*,         mark options={solid, fill=cNoCal}},
  sweep p4/.style   ={color=cP4,    mark=square*,    densely dashed, mark options={solid, fill=cP4}},
  sweep p8/.style   ={color=cP8,    mark=triangle*,  mark options={solid, fill=cP8, scale=1.25}},
  sweep p16/.style  ={color=cP16,   mark=diamond*,   densely dotted, mark options={solid, fill=cP16, scale=1.3}},
}
\begin{tikzpicture}
\begin{groupplot}[
  group style={group size=1 by 2, vertical sep=8pt,
               xticklabels at=edge bottom, xlabels at=edge bottom},
  pgf2 base,
  sweep table,
  height=0.5\linewidth,
  xmin=\chimin, xmax=\chimax, enlarge x limits=false,
  ymax = 7,
  xtick={7,9,...,27}, xticklabels={}, xticklabel={\pgfmathprintnumber{\tick}},
  minor xtick={6,8,...,26},
  xlabel={Detection threshold $\chi$ (dB)},
  legend columns=1,
]
  \nextgroupplot[
    ymin=0,                                
    enlarge y limits={upper, value=0.06},
    ylabel={OSPA$_{\mathrm{pos}}$ (m)},
    legend to name=sweepLegend,
  ]
  \addplot+[sweep nocal, sweep variant={(no cal)}{4}] table[x index=3, y index=\sweepmetric] {pgf2/threshold_sweep_sim.csv};
  \addlegendentry{No Cal., P = 4}
  \addplot+[sweep p4,  sweep variant={(cal)}{4}]  table[x index=3, y index=\sweepmetric] {pgf2/threshold_sweep_sim.csv};
  \addlegendentry{Cal., P = 4}
  \addplot+[sweep p16, sweep variant={(cal)}{16}] table[x index=3, y index=\sweepmetric] {pgf2/threshold_sweep_sim.csv};
  \addlegendentry{Cal., P = 16}

  \nextgroupplot[
    ymin=0, enlarge y limits={upper, value=0.08},
    ytick distance=1,
    ylabel={$\hat{K}$},
  ]
  \addplot+[sweep nocal, sweep variant={(no cal)}{4}] table[x index=3, y index=\sweepkhat] {pgf2/threshold_sweep_sim.csv};
  \addplot+[sweep p4,  sweep variant={(cal)}{4}]  table[x index=3, y index=\sweepkhat] {pgf2/threshold_sweep_sim.csv};
  \addplot+[sweep p16, sweep variant={(cal)}{16}] table[x index=3, y index=\sweepkhat] {pgf2/threshold_sweep_sim.csv};
  \draw[black, dashdotted, line width=1.0pt]
    ({rel axis cs:0,0} |- {axis cs:0,2}) -- ({rel axis cs:1,0} |- {axis cs:0,2})
    node[pos=1, anchor=south east, font=\footnotesize, inner sep=2pt, text=black] {true $K$};
\end{groupplot}
\node[anchor=north east, inner sep=0pt]
  at ([xshift=-3pt, yshift=-3pt]group c1r2.north east) {\ref*{sweepLegend}};
\end{tikzpicture}
  \caption{Simulated influence of the detection threshold $\chi$ on
    \gls{ospa} of the position estimates (top) and average number of estimated
    components $\hat{K}$ (bottom).}
\label{fig:threshold_sweep}
\end{figure}

  \textbf{Sensing performance over detection threshold:}
  Second, we evaluate the \gls{rmse} of the \gls{ospa} metric for source position estimation as a function of the detection threshold $\chi$. We compute the estimated position of source component $k$ as $c\hat\tau_k[\sin\hat\varphi_k,\ \cos\hat\varphi_k]^\mathsf{T}$ to jointly account for delay and angle estimates. We generate $3$ different fixed hardware deviations with medium severity of $\sigma_w =0.15$ and $\sigma_{\Delta\rho}=0.1 \lambda_c$. We consider $5$ different scenarios with $2$ components each, where $c\tau_k \sim \mathcal{U}(1\,\mathrm{m}, 10\,\mathrm{m})$ and $\varphi_k \sim \mathcal{U}(-60^\circ, 60^\circ)$ with a minimum separation of $c\Delta\tau = 1$~m and $\Delta\varphi = 10^\circ$, and fixed \glspl{snr} of $30$ dB and $15$ dB, where the closer component has the higher \gls{snr}. We combine the hardware deviations and the scenarios to get $15$ different simulation runs per detection threshold. In Fig. \ref{fig:threshold_sweep}, we compare self-calibration with $P=4$ and $P=16$ antenna elements against the uncalibrated variant with $P=4$. We also plot the average number of estimated components. The algorithm performs well within a certain range of detection thresholds. At very low thresholds, spurious components may be mistakenly detected. Conversely, at very high thresholds, true components may be incorrectly pruned. Increasing the number of antenna elements shifts the \gls{ospa} curve slightly toward higher detection thresholds due to the increased array processing gain. Moreover, a larger number of antenna elements improves the position estimation accuracy. Despite the hardware deviations, the position \gls{ospa} reaches approximately $0.6$ m with $P=4$ elements and $0.1$ m with $P=16$ elements.
  
\label{subsec:eval_random}
\subsection{Measurement Data}
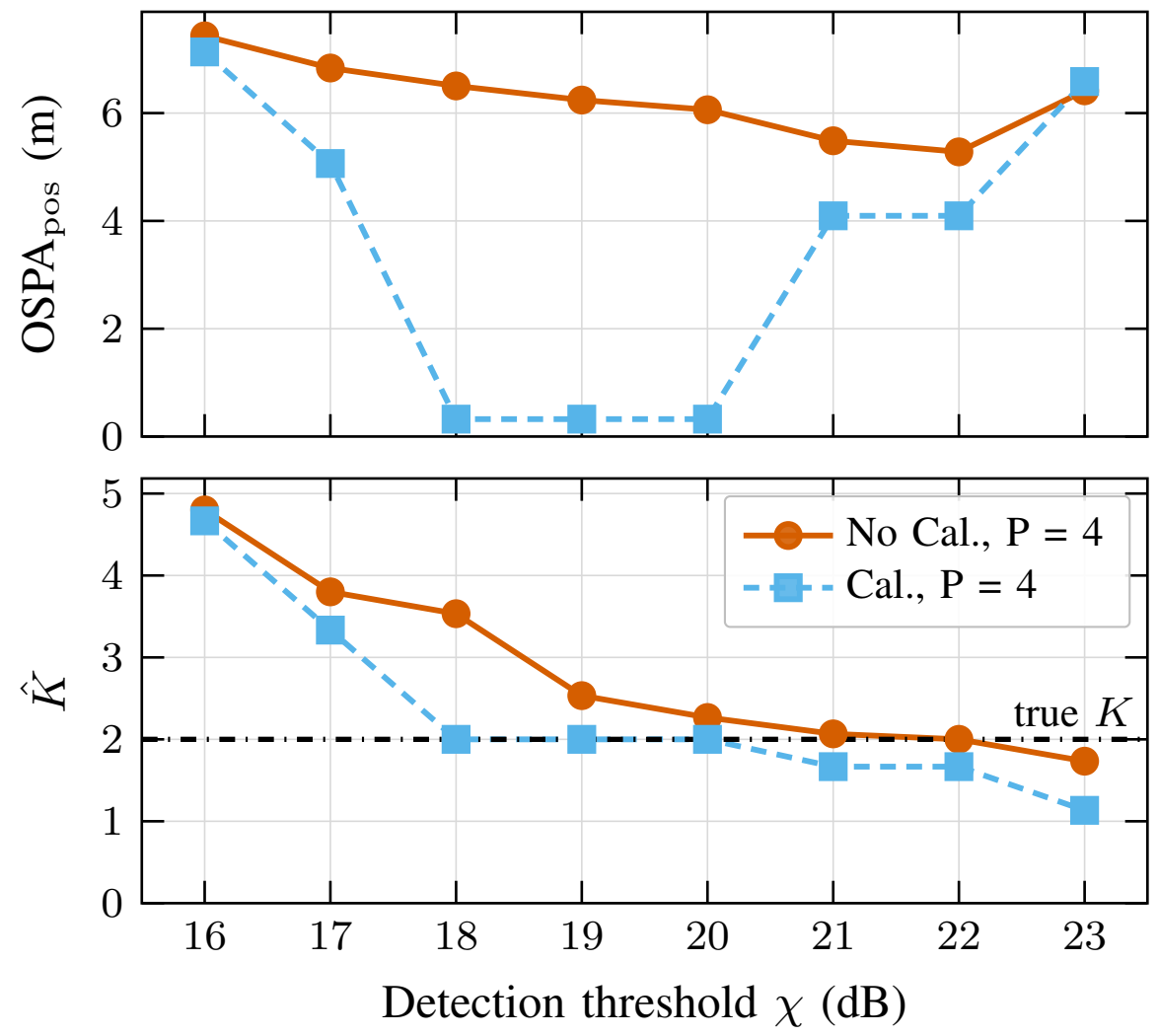
\begin{figure}[]
  \centering
%
\usepgfplotslibrary{groupplots}%
\def\measchimin{15.5}\def\measchimax{23.5}%
\providecolor{cP4}{HTML}{56B4E9}
\def\pgftwomeaskeep#1#2{%
  \def\pgftwomeastest##1#1##2\pgftwomeasend{%
    \ifx\relax##2\relax\def\pgfmathresult{nan}\fi}%
  \pgftwomeastest#2#1\pgftwomeasend}%
\pgfplotsset{
  meas table/.style={table/col sep=comma, table/header=false, table/skip first n=1},
  meas variant/.style={
    x filter/.code={%
      \edef\pgftwomeasdo{\noexpand\pgftwomeaskeep{#1}{\thisrowno{0}}}%
      \pgftwomeasdo},
  },
  meas nocal/.style={color=cNoCal, mark=*,       mark options={solid, fill=cNoCal}},
  meas cal/.style  ={color=cP4,    mark=square*, densely dashed, mark options={solid, fill=cP4}},
}
\begin{tikzpicture}
\begin{groupplot}[
  group style={group size=1 by 2, vertical sep=8pt,
               xticklabels at=edge bottom, xlabels at=edge bottom},
  pgf2 base,
  meas table,
  height=0.5\linewidth,
  xmin=\measchimin, xmax=\measchimax, enlarge x limits=false,
  xtick={16,17,...,23}, xticklabels={}, xticklabel={\pgfmathprintnumber{\tick}},
  minor xtick={},
  xlabel={Detection threshold $\chi$ (dB)},
  legend columns=1,
]
  \nextgroupplot[
    ymin=0,                                
    enlarge y limits={upper, value=0.06},
    ylabel={OSPA$_{\mathrm{pos}}$ (m)},
    legend to name=measLegend,
  ]
  \addplot+[meas nocal, meas variant={(no cal)}] table[x index=1, y index=2] {pgf2/threshold_sweep_meas.csv};
  \addlegendentry{No Cal., P = 4}
  \addplot+[meas cal,   meas variant={(cal)}]    table[x index=1, y index=2] {pgf2/threshold_sweep_meas.csv};
  \addlegendentry{Cal., P = 4}

  \nextgroupplot[
    ymin=0, enlarge y limits={upper, value=0.08},
    ytick distance=1,
    ylabel={$\hat{K}$},
  ]
  \addplot+[meas nocal, meas variant={(no cal)}] table[x index=1, y index=3] {pgf2/threshold_sweep_meas.csv};
  \addplot+[meas cal,   meas variant={(cal)}]    table[x index=1, y index=3] {pgf2/threshold_sweep_meas.csv};
  \draw[black, dashdotted, line width=1.0pt]
    ({rel axis cs:0,0} |- {axis cs:0,2}) -- ({rel axis cs:1,0} |- {axis cs:0,2})
    node[pos=1, anchor=south east, font=\footnotesize, inner sep=2pt, text=black] {true $K$};
\end{groupplot}
\node[anchor=north east, inner sep=0pt]
  at ([xshift=-3pt, yshift=-3pt]group c1r2.north east) {\ref*{measLegend}};
\end{tikzpicture}
  \caption{Measured influence of the detection threshold $\chi$ on
    \gls{ospa} of the position estimates (top) and average number of estimated
    components $\hat{K}$ (bottom).}
\label{fig:threshold_sweep_meas}
\end{figure}

The measurement data were collected with an omnidirectional antenna as transmitter and a 4$\times$1 analog phased-array antenna as receiver in an indoor laboratory. At three positions, we measured the \gls{los} component and one \gls{mpc}, blocking other paths with absorbers, while the ground truth was obtained via motion capture and geometry. We measured five snapshots per position. In each snapshot, the analog phased array was steered to $15$ beam directions, and the resulting beamspace data were transformed to elementspace as in \cite{SayGuvDoeSemGen:TWC2024} before applying our algorithm. The \gls{rmse} of the \gls{ospa} metric is computed over all snapshots and positions. In Fig. \ref{fig:threshold_sweep_meas}, we plot the sensing performance as a function of the detection threshold, together with the estimated number of components, comparing the algorithm with and without self-calibration. The results exhibit a trend similar to that observed with the synthetic data in the detection-threshold sweep. There is an optimal threshold range where we detect the correct number of components and reach a position \gls{ospa} of around $0.2$~m. Although the measurements were taken with a pre-calibrated antenna, the algorithm without calibration exhibits significantly worse sensing performance. This indicates residual model mismatch, such as hardware impairments, which the proposed algorithm was able to mitigate.

\section{Conclusion}
This study presents a Bayesian wideband parametric channel estimation algorithm with integrated self-calibration of complex weights and per-antenna position deviations. An update rule is derived for the antenna element position vector from the marginal likelihood within a Bayesian framework. Moreover, the \gls{hcrlb} is derived as a performance benchmark for our simulations. The algorithm is evaluated on synthetic data and measured data captured by an analog phased-array frontend.

The results demonstrate the benefits of joint calibration compared with gain-only, position-only, and no calibration across the considered impairment scenarios. The \gls{hcrlb} benchmark is approached over a wide range, except under severe hardware impairments. Furthermore, the results show the importance of selecting an appropriate detection threshold for the proposed algorithm. Substantial improvements are demonstrated on single-snapshot measurements from a pre-calibrated phased array, where self-calibration reduced the position OSPA from about 6~m to 0.2~m. These findings highlight the relevance of integrating self-calibration into parametric channel estimation for practical measurement setups. Future work will focus on the fundamental limits of self-calibration, particularly the identifiability of channel and calibration parameters. Additionally, the most efficient physical calibration models need to be identified for practical measurement setups, as they will play a key role in future 6G integrated sensing systems.

   \section*{Appendix: Derivation of HCRLB Benchmark}
   The \glspl{doa} and delays $\boldsymbol{\theta}$ are the deterministic parameters of interest. The path amplitudes $\boldsymbol{\alpha}$ are treated as deterministic nuisance parameters, fixed by their \glspl{snr}, and the benchmark assumes a known number of components $K$. The complex weight deviations $\bm{w}\sim\mathcal{CN}(\bm{1},\sigma_w^2\bm{I}_P)$ are
random, while the position deviations $\Delta\boldsymbol{\rho}$, deterministic in the signal model but drawn synthetically, are assigned the prior $\Delta\boldsymbol{\rho}\sim\mathcal{N}(\bm{0},\sigma_{\Delta\rho}^2\bm{I}_P)$.
This gives $\boldsymbol{\nu}_{\mathrm d} = [\Re\{\boldsymbol{\alpha}\}^{\mathsf T}, \Im\{\boldsymbol{\alpha}\}^{\mathsf T}]^{\mathsf T}$, $\boldsymbol{\nu}_{\mathrm r} = [\Re\{\bm{w}\}^{\mathsf T}, \Im\{\bm{w}\}^{\mathsf T}, \Delta\boldsymbol{\rho}^{\mathsf T}]^{\mathsf T}$, and the hybrid parameter vector
$\boldsymbol{\psi} = [\boldsymbol{\theta}^{\mathsf T}, \boldsymbol{\nu}_{\mathrm d}^{\mathsf T}, \boldsymbol{\nu}_{\mathrm r}^{\mathsf T}]^{\mathsf T}$. Following~\cite[Eq.~(13)]{MesWSAMP2006}, the hybrid \gls{fim} is defined as
\begin{align}
  \bm{J}
  &= -\mathbb{E}_{\bm{y},\boldsymbol{\nu}_\mathrm{r}}\!\left[
     \frac{\partial^2 \ln p(\bm{y},\boldsymbol{\nu}_\mathrm{r};\boldsymbol{\theta},\boldsymbol{\nu}_\mathrm{d})}
          {\partial\boldsymbol{\psi}\,\partial\boldsymbol{\psi}^{\mathsf T}}\right]
     \label{eq:hfim_def}\\
  &= -\mathbb{E}_{\bm{y},\boldsymbol{\nu}_\mathrm{r}}\!\left[
     \frac{\partial^2 \ln p(\bm{y}|\boldsymbol{\psi})}
          {\partial\boldsymbol{\psi}\,\partial\boldsymbol{\psi}^{\mathsf T}}\right]
     -\mathbb{E}_{\bm{y},\boldsymbol{\nu}_\mathrm{r}}\!\left[
     \frac{\partial^2 \ln p(\boldsymbol{\nu}_\mathrm{r})}
          {\partial\boldsymbol{\psi}\,\partial\boldsymbol{\psi}^{\mathsf T}}\right]
     \label{eq:hfim_split}\\
  &= \mathbb{E}_{\boldsymbol{\nu}_\mathrm{r}}\!\Bigg[\underbrace{-\mathbb{E}_{\bm{y}|\boldsymbol{\psi}}\!\left[
     \frac{\partial^2 \ln p(\bm{y}|\boldsymbol{\psi})}
          {\partial\boldsymbol{\psi}\,\partial\boldsymbol{\psi}^{\mathsf T}}\right]}_{\bm{J}_\mathrm{D}(\boldsymbol{\psi})}\Bigg]
     \;\underbrace{-\;\mathbb{E}_{\boldsymbol{\nu}_\mathrm{r}}\!\left[
     \frac{\partial^2 \ln p(\boldsymbol{\nu}_\mathrm{r})}
          {\partial\boldsymbol{\psi}\,\partial\boldsymbol{\psi}^{\mathsf T}}\right]}_{\bm{J}_\mathrm{P}},
     \label{eq:hfim_nested}
\end{align}
where~\eqref{eq:hfim_split} follows from the factorization
$p(\bm{y},\boldsymbol{\nu}_\mathrm{r};\boldsymbol{\theta},\boldsymbol{\nu}_\mathrm{d})
 = p(\bm{y}|\boldsymbol{\psi})\,p(\boldsymbol{\nu}_\mathrm{r})$ and
\eqref{eq:hfim_nested} from the law of total expectation
$\mathbb{E}_{\bm{y},\boldsymbol{\nu}_\mathrm{r}}[\cdot]
 = \mathbb{E}_{\boldsymbol{\nu}_\mathrm{r}}\big[\mathbb{E}_{\bm{y}|\boldsymbol{\psi}}[\cdot]\big]$.
$\bm{J}_\mathrm{D}(\boldsymbol{\psi})$ is the data \gls{fim}
conditioned on $\boldsymbol{\nu}_\mathrm{r}$ for fixed $\boldsymbol{\theta}$ and
$\boldsymbol{\nu}_\mathrm{d}$, and $\bm{J}_\mathrm{P}$ is the Bayesian \gls{fim}.
Since $\bm{y}\sim\mathcal{CN}\big(\bm\mu(\bm\psi),\lambda^{-1}\bm{I}\big)$
with $\bm\mu(\bm\psi) = \bm D(\bm w)\,\bm A(\bm\theta,\Delta\bm\rho)\,\bm\alpha$
and the true noise precision $\lambda$ assumed to be known, the data \gls{fim}
follows as
\begin{equation}
  [\bm J_{\mathrm D}(\bm\psi)]_{\ell\ell'}
  = 2\lambda\,
  \Re\!\left\{
    \left(\frac{\partial\bm\mu}{\partial\psi_\ell}\right)^{\mathrm H}
    \frac{\partial\bm\mu}{\partial\psi_{\ell'}}
  \right\},
  \label{eq:nb_data_fim}
\end{equation}
while the Gaussian priors on $\bm w$ and $\Delta\bm\rho$ yield the constant
diagonal Bayesian \gls{fim}
\begin{equation}
  \bm J_{\mathrm P} = \mathrm{diag}\Big(
    \bm 0_{4K}^{\mathsf T},\;
    \tfrac{2}{\sigma_w^2}\bm 1_{2P}^{\mathsf T},\;
    \tfrac{1}{\sigma_{\Delta\rho}^2}\bm 1_{P}^{\mathsf T}
  \Big),
\end{equation}
with zero entries for $\bm\theta$ and $\bm\nu_\mathrm{d}$. Unlike~\cite{MesWSAMP2006}, we omit the outer expectation of the data \gls{fim}
over $\bm\nu_\mathrm{r}$ in~\eqref{eq:hfim_nested}, and instead evaluate
$\bm J^{(i)} = \bm J_\mathrm{D}(\bm\psi^{(i)}) + \bm J_\mathrm{P}$
at the true $\bm\theta$, $\bm\nu_\mathrm{d}$ and the $i$-th impairment realization
$\bm\nu_\mathrm{r}^{(i)}$, as in the simulations.
    Partitioning $\bm J^{(i)}$ according to $\bm\theta$ and
$\bm\nu = [\bm\nu_{\mathrm d}^{\mathsf T}, \bm\nu_{\mathrm r}^{\mathsf T}]^{\mathsf T}$,
\begin{equation}
  \bm J^{(i)} = \begin{bmatrix}
    \bm J_{\bm\theta\bm\theta}^{(i)} & \bm J_{\bm\theta\bm\nu}^{(i)} \\
    \bm J_{\bm\nu\bm\theta}^{(i)} & \bm J_{\bm\nu\bm\nu}^{(i)}
  \end{bmatrix},
\end{equation}
the nuisance parameters are eliminated via the Schur complement, yielding
\begin{equation}
  \bm C^{(i)}
  = \big(\bm J_{\bm\theta\bm\theta}^{(i)}
  - \bm J_{\bm\theta\bm\nu}^{(i)}\,
    (\bm J_{\bm\nu\bm\nu}^{(i)})^{-1}\,
    \bm J_{\bm\nu\bm\theta}^{(i)}\big)^{-1}.
\end{equation}
The average of $\bm C^{(i)}$ over the $I$ realizations then approximates its expectation over $\bm\nu_\mathrm{r}$, which, together with the average over the $K$ components, yields the HCRLB benchmarks
    \begin{equation}
      \begin{aligned}
        \mathrm{\gls{hcrlb}}_{\varphi} &= \sqrt{\frac{1}{IK}\sum_{i=1}^{I}\sum_{k=1}^{K}
          \big[\bm{C}^{(i)}\big]_{k,k}}, \\
        \mathrm{\gls{hcrlb}}_{\tau} &= \sqrt{\frac{1}{IK}\sum_{i=1}^{I}\sum_{k=1}^{K}
          \big[\bm{C}^{(i)}\big]_{K+k,K+k}}.
      \end{aligned}
      \label{eq:hcrb_avg}
    \end{equation}
  For OSPA order $q=2$ and $\hat{K}=K$, the \gls{ospa} metric \cite{SchVoVo:TSP2008} has the same \gls{rmse} structure as \eqref{eq:hcrb_avg}. We deliberately include runs with $\hat{K}\neq K$, so that
    a gap to the benchmark reveals cardinality errors penalized by our cutoff parameters.

\section*{Acknowledgment}
This work was supported in part by the DragonGEM project, funded by the Austrian Research Promotion Agency (FFG). The authors used Claude (Anthropic) for language
editing.

\bibliographystyle{IEEEtran}
\bibliography{IEEEabrv,references_polished_updated.bib}

\end{document}